\documentstyle[aps,prb,twocolumn]{revtex}
\newcommand{\ccc}{{\c C}a{\u g}{\i}n}
\begin{document}
\wideabs{
\title{First Principles Force Field for Metallic Tantalum}

\author{Alejandro Strachan $^1$, Tahir \ccc $^1$, O{\u g}uz G{\"u}lseren $^2$
Sonali Mukherjee $^{2,3}$, Ronald E. Cohen $^{2,3}$
and William A. Goddard III $^1$ *}

\address{$^1$ Materials and Process Simulation Center,
Beckman Institute (139-74)\\
California Institute of Technology, Pasadena, California 91125.\\
$^2$ Geophysical Laboratory and Center for High Pressure Research\\ 
Carnegie Institution of Washington,\\
5251~Broad~Branch~Road,~NW,~Washington,~DC~20015.\\
$^3$ Seismological Laboratory, California Institute of Technology, Pasadena, CA
91125}

\date{\today}

\maketitle

\begin{abstract}

We propose a general strategy to develop accurate Force Fields (FF) for
metallic systems derived from {\it ab initio} quantum mechanical (QM) calculations;
we illustrate this approach for tantalum.
As input data to the FF we use the linearized augmented plane wave method (LAPW) with
the generalized gradient approximation (GGA) to calculate:

(i) the zero temperature equation of state (EOS) of Ta for bcc, fcc, 
and hcp crystal structures for pressures up to $\sim$ 500 GPa. 

(ii) Elastic constants.

(iii) We use a mixed-basis pseudopotential code to calculate volume
relaxed vacancy formation energy also as a function of pressure.

In developing the Ta FF we also use previous QM calculations of:

(iv) the equation of state for the A15 structure.

(v) the surface energy bcc (100).

(vi) energetics for shear twinning of the bcc crystal.

We find that with appropriate parameters an embedded atom model force field (denoted as qEAM FF) 
is able to reproduce all this QM data. Thus, the {\it same} FF describes 
with good accuracy the bcc, fcc, hcp and A15 phases of Ta for pressures from $\sim -10$ GPa to $\sim 500$ 
GPa, while also describing the vacancy, surface energy, and shear transformations. 
The ability of this single FF to describe such a range of systems with a variety of coordinations 
suggests that it would be accurate for describing defects such as dislocations, grain boundaries, etc.

We illustrate the use of the qEAM FF with molecular dynamics to calculate such finite temperature 
properties as the melting curve up to $300$ GPa; we obtain a zero pressure melting temperature of
$T_{melt}=3150 \pm 50$ K in good agreement with experiment ($3213-3287$ K). We also report on the 
thermal expansion of Ta in a wide temperature range; our calculated thermal expansivity agrees well 
with experimental data.

\end{abstract}
}
{64.30.+t, 62.20.Dc, 64.70.-p, 64.70.Dv, 65.40.-b}

\section{Introduction}

Despite decades of experimental and theoretical research on the mechanical properties 
of materials many questions remain open, particularly in the relation between 
atomistic processes (involving dislocations, grain boundaries, cracks, their mobility and
interactions thereof) and the macroscopic 
behavior (plastic deformation, failure, etc.). Macroscopic plasticity and failure are well 
characterized experimentally and described using a variety of mesoscale and 
macroscale models with parameters obtained empirically. These models and their parameters should ultimately 
be derivable in terms of the fundamental physics of atomic interactions as described by quantum 
mechanics (QM). Unfortunately despite the enormous progress in {\it ab initio} QM, such calculations are 
too computationally demanding to study directly most processes relevant to plasticity and failure. 
To do so would require that millions of atoms be described for nanoseconds or longer, calculations that 
are impossible to consider today. In order to bridge this gap between atomic interactions and the mechanical
properties of macroscopic 
systems we use first principles QM data to derive a FF with which energies and forces 
can be calculated in a computationally efficient way given only the atomic positions. This allows us 
to use classical molecular 
dynamics (MD) to simulate the various atomistic processes governing the mechanical and thermodynamical 
properties of materials. Since the FF is derived by fitting a wide range of QM data, we expect to obtain
an accurate description of the atomic interactions which, with molecular dynamics, can provide insight and 
constitutive equations to be used in the mesoscale and macroscale models of plasticity and failure.
We illustrate this procedure here for Ta. Ta is chosen because it exhibits only a single crystalline phase 
over the interesting range of temperatures and pressures. This makes the validation of theoretical
predictions against experimental data under a wide range of conditions more clear.

Several authors have used {\it ab initio} methods to study various mechanical properties
of Ta. S{\"o}derlind and Moriarty used the full potential linear muffin-tin 
orbital method within the GGA approximation and with
spin orbit interactions to calculate different zero temperature properties of 
of Ta, including the EOS of different crystalline phases, elastic constants, and
shear strength. \cite{soderlind98} Vacancy formation migration energies have also been 
calculated from first principles. \cite{soderlind00,satta99} Recently, Ismail-Beigi 
and Arias \cite{sohrab00} calculated {\it ab initio} dislocation core energy and structures 
for Ta and Mo.

On the other hand, several FF have been developed for bcc metals. The first principles 
based multi-ion analytic model generalized pseudopotential (MGPT) force field has been 
used to calculate several mechanical
and thermodynamical properties of various metals, \cite{mori_momelt,mori_modisl,mori_ta} including Ta.
One of the most popular many-body force fields for metals is the 
Embedded Atom Model (EAM), proposed by Daw and Baskes in 1984. \cite{daw84}
Most EAM FF have been based on experimental data regarding structures at and near equilibrium
[for bcc metals, see for example Ref. \onlinecite{chanta96,johnson89,guellil92}]. One of the main
advantages of EAM force fields is that they are computationally very efficient which
allows MD simulations of large systems for long times. The modified embedded atom model
(MEAM), which includes angular dependence of the electronic density, was developed by Baskes and 
coworkers \cite{MEAM} and applied to a variety of materials including bcc transition metals 
\cite{MEAM_metal}.

In this paper we propose a strategy to derive accurate many body FF based on
QM calculations; these FF can be used to simulate the dynamics of systems with millions
of atoms. We use accurate QM (LAPW GGA method) to calculate
various mechanical properties of Ta which require a small number of atoms:
namely the zero temperature EOS for bcc, fcc and hcp
crystalline phases, and elastic constants. We also use a mixed-basis pseudopotential method to
calculate vacancy formation energies. 
We find that we can describe all this first principles data using a classical many body
EAM FF (named qEAM FF) with good overall accuracy.
We then illustrate the use of the qEAM FF with MD
simulations to study various properties as a function of pressure and temperature;
such as the melting curve of Ta up to pressures of $\sim 300$ GPa, and thermal 
expansivity.

This paper is organized as follows. In Section II we present first
principles results for the equation of state for bcc, fcc and hcp phases,
elastic constants and volume relaxed vacancy formation energy and enthalpy
for a wide pressure range. In Section III we develop the qEAM FF based on
{\it ab initio} results. In Section IV we calculate the thermal expansivity of Ta using
the qEAM FF and {\it ab initio} calculations. Section V presents the calculation of the melting curve
of Ta for pressures up to $300$ GPa using molecular dynamics. Finally,
in section VI, conclusions are drawn.

\section{Quantum mechanics results}

We computed the static equation of state of Ta for different crystalline phases using 
the linearized augmented 
plane wave (LAPW) method. \cite{weikrakauer,singh} LAPW is an 
all-electron method, with no essential shape approximations for the charge 
density or potential, and is easily converged. The $5p$, $4f$, $5d$ and 
$6s$ states were treated as band states, and the deeper states were 
treated as soft core electrons. Here we used the Perdew, Burke, and Ernzerhof (PBE) 
implementation of the generalized gradient 
approximation \cite{PBE} for the exchange-correlation potential. 
A 16x16x16 special k-point mesh\cite{monpack} was used, giving 140 k-points 
within the irreducible Brillouin zone of the bcc lattice. Tests 
demonstrated convergence with this mesh. The convergence 
parameter RK$_{max}$ was 9 giving about 1800 planewaves and 200 basis functions
per atom at zero pressure. 

Total energies were computed for bcc, fcc and hcp phases at 20 volumes. 
For the fcc and hcp phases, 12x12x12 and 16x16x12 k-point meshes 
were used for Brillouin zone integrations giving 182 and 180 k-points within the irreducible zone 
respectively. For the hcp phase, the ideal c/a ratio was used, and 
at two different volumes also c/a was optimized. We found that the change 
in the energy due to the this optimization is less than 40~meV/atom around 
zero pressure and decreases by pressure. 
We have tabulated the {\it ab initio} energy-volume data for the different phases and have made them 
available as supplementary material, see Ref. \onlinecite{supp_mat}. We have fitted our
energy-volume data to Rose's universal equation of state; \cite{vinet} the obtained
zero pressure volume ($V_0$), zero temperature bulk modulus ($B_T$), and its derivative
with respect to pressure ($B_T^{'}$) are shown in Table \ref{table_bcc}. We also show
in Table \ref{table_bcc} the results obtained by S{\"o}derlind and Moriarty using full potential 
linear muffin-tin orbital method within the GGA approximation and with spin orbit interactions (denoted as
FP LMTO GGA SC), and room temperature experimental values by Cynn and Yoo. \cite{cynn}
Our LAPW calculations of the bcc equation of state agree well with the experimental values 
and previous theoretical calculations.

Static elastic constants [c$_s$=(c$_{11}$-c$_{12}$)/2 and c$_{44}$] were obtained from 
strain energies by straining the bcc cell with volume conserving tetragonal and orthorhombic.
We calculate c$_s$ using tetragonal strain of the cubic bcc lattice:
\begin{eqnarray}
\mathbf{a} & = & a ( 1+\epsilon ,          0 , 0               ),\nonumber \\
\mathbf{b} & = & a ( 0          , 1+\epsilon , 0               ),\nonumber \\
\mathbf{c} & = & a ( 0          ,          0 , 1/(1+\epsilon)^2),
\end{eqnarray}
where $a$  is the cubic lattice constant of the system, $\mathbf{a}$, $\mathbf{b}$, 
and $\mathbf{c}$ are the lattice vectors and $\epsilon$ is the strain.
$c_s$ is related to the quadratic term of the strain energy:
\begin{equation}
E(\epsilon) = E(\epsilon=0) + 6  V(\epsilon=0)  c_s  \epsilon^2 + O(\epsilon^3),
\end{equation}
where $E(\epsilon=0)$ is the energy of the unstrained system and $V(\epsilon=0)$ is its
volume.
Similarly c$_{44}$ is obtained from the orthorhombic strain:
\begin{eqnarray}
\mathbf{a} & = & a ( 1 , \epsilon , 0),\nonumber \\
\mathbf{b} & = & a ( \epsilon, 1 , 0),\nonumber \\
\mathbf{c} & = & a ( 0 , 0 , 1/(1 - {\epsilon}^2)).
\end{eqnarray}
the shear constant $c_{44}$ is obtained from:
\begin{equation}
E(\epsilon) = E(\epsilon=0) + 2  V(\epsilon=0)  c_{44}  \epsilon^2 + O(\epsilon^3) .
\end{equation}

The convergence of strain energies with respect to the Brillouin zone integration was 
carefully checked; we used 16x16x16 k-points meshes in the full 
Brillouin zone giving 344 and 612 k-points within the irreducible Brillouin zone of tetragonal and 
orthorhombic lattice respectively. $c_s$ and $B_T=(c_{11}+2c_{12})/3$ were used to calculate 
$c_{11}$ and $c_{12}$;
the resulting zero pressure and zero temperature elastic constants are
shown in Table \ref{table_bcc}. 
The elastic constants $c_s$ and $c_{44}$ as functions of pressure are available as
supplementary material\cite{supp_mat}, see also Ref. \onlinecite{oguz2001}. The 
zero pressure values and initial slopes are in good agreement with the 
experimental data of Katahara et al.\cite{katahara}
We find that c$_{44}$ shows a change in behavior at $\sim$150 GPa \cite{oguz2001} which is probably due to the 
electronic transition also evident in the equation of state.\cite{cohen_thermal} 
The band structure and density of state show a major reconfiguration
with pressure.\cite{cohen_thermal}
Our results indicate that the elastic constants can be much more sensitive to 
changes in the occupied states below the Fermi surface than the equation of state, where changes 
were much more subtle.

The vacancy formation energy was determined \cite{sonali} using supercells 
of 16 and 54 atoms and a mixed-basis pseudopotential (MB-PS) code.\cite{oguz} Using 
the GGA approximation, \cite{pw91} the zero pressure volume relaxed vacancy formation 
energy was determined to be 3.25 eV for a 54 atom supercell, and 3.26 eV 
for a 16 atom supercell, indicating convergence. The effects of structural
relaxation on vacancy formation energy are discussed in Ref. \onlinecite{sonali}.   
A comparison with recent {\it ab initio} and experimental results will be presented in
the next section. The {\it ab initio} data used to calculate vacancy formation energies
is also available as supplementary material.\cite{supp_mat}

\section{\lowercase{q}EAM force field}

In order to predict mechanical properties of materials and processes like
phase diagrams, dislocations structures and mobility, mechanical 
failure, etc. it is important to have accurate classical force fields
to describe the atomic interactions. With MD simulations it is then
possible to study large systems (millions of atoms) for relatively
long times (ns).

One of the most popular many-body force fields for metals is the 
EAM, proposed by Daw and Baskes in 1984. \cite{daw84}
This approach is computationally efficient and has been used successfully
for numerous applications, like calculation of thermodynamic functions,
liquid metals, defects, grain boundary structure, fracture, etc, see for
example Ref. \onlinecite{guellil92,daw93}.

The EAM implementation that we have used is based on the one proposed
by Chantasiriwan and Milstein. \cite{chanta96} This from of EAM was chosen because it can
describe third order elastic constants correctly \cite{chanta96} thus being useful in a
wide range of strains. The energy of a given atomic 
configuration with atom positions $\{r_i\}$ is given by:

\begin{equation}
U\{r_i\} = \sum_i^N F(\rho_i) + \sum_{i<j} \phi(r_{ij}),
\end{equation}
with 
\begin{equation}
\rho_i = \sum_{j\neq i} f(r_{ij})
\end{equation}
where $F(\rho)$ is the embedding energy, $\rho_i$ is the total ``electronic density''
on the atomic site $i$, $f(r)$ is the electron density function,
$\phi(r)$ is a two-body term and $r_{ij} = |r_i - r_j|$.

Following Ref. \onlinecite{chanta96} we took the two-body term as:
\begin{equation}
\phi(r) = \left\{ \begin{array}{ll}
	(r-r_m)^4 \sum_{i=0}^7 b_i r^i  & \mbox{if $r<r_m$} \\
	0				& \mbox{otherwise,}
		\end{array}
\right.
\end{equation}	
the factor $(r-r_m)^4$ ensures that the two body term and its first three derivatives with respect
to $r$ vanish at the the cut-off radius ($r_m$). The optimized form of the two body term shows
short range repulsion and longer range attraction.

The ``electron density'' is:
\begin{equation}
f(r) = \frac{ 1+a_1 \cos(\frac{\alpha r}{V^{1/3}}) +
a_2 \sin(\frac{\alpha r}{V^{1/3}}) }{r^{\beta}},
\end{equation}
where $V$ is the volume per atom of the system. The parameters of the ``electron 
density'' are volume dependent, but structure independent. The importance of the
oscillatory behavior of the ``electron density function'' in embedded atom model-like
force fields is related to their ability to correctly account anharmonicities. \cite{chanta96}

Finally the embedding energy as a function of the electronic density is obtained from 
the reference bcc structure: \cite{chanta96}
\begin{equation}
F(\rho) = U_{Rose}(V) - \sum_{i<j} \phi(r_{ij}),
\end{equation}
where the sum is made for the perfect bcc structure and $U_{Rose}(V)$ is Rose's 
universal  equation of state: \cite{vinet}
\begin{equation}
U_{Rose}(V) = -E_{coh}(1+a^* + f_3 a^{*3} +f_4 a^{*4}) e^{-a^*}
\end{equation}
where $a^* = (a - a_0)/a_0 \lambda$ and $\lambda = (E_{coh}/9 V_0
B_T)^{1/2}$.

We we optimized the parameters in the qEAM FF using the following data:

(a) Zero temperature energy-volume and pressure-volume curves for 
different crystal structures (bcc, fcc and A15) in a wide pressure range, 
from $\sim -10$ GPa to $\sim 500$ GPa. For the bcc and fcc structures
the data from Section II was used, while for A15 we used the results
obtained by S{\"o}derlind and Moriarty.\cite{soderlind98}

(b) Zero temperature, zero pressure elastic coefficients, shown in section II.

(c) Vacancy formation energy at zero pressure.

(d) Energetic of homogeneously sheared bcc crystal, from Ref. \onlinecite{soderlind98}.

(c) Unrelaxed (100) surface energy from Ref. \onlinecite{surf_qm}.

We fit the parameters entering the qEAM FF energy expression
to the training set using an optimization algorithm based on simulated annealing.
We define an error function of the form:
\begin{equation}
C = \sum_i C_i \frac{(Q^{qEAM}_i - Q^{target}_i)^2}{(Q^{target}_i)^2},
\end{equation}
where the sum runs over the different quantities to be fitted, $Q^{qEAM}_i$
is the value given by the qEAM FF for quantity $i$, $Q^{target}_i$ is the
target quantity and $C_i$ are weight factors. The EOS information of the different
phases account for $\sim78 \%$ of the total cost; elastic constants and vacancy
formation energy account for $\sim 9\%$ each one; surface energy represents
$\sim 3\%$ of the total cost and energy of sheared bcc crystal represent
$\sim 1\%$.

In Tables \ref{table_ff1} and \ref{table_ff2} we show the optimized qEAM parameters.

A detailed comparison between the qEAM FF and the data it was fitted to
is shown in the following subsections, together with other {\it ab initio} and
experimental results. We also show the calculation of related quantities obtained using the
qEAM FF via MD simulations.

\subsection{Equation of state and elastic constants}

The most important quantities used to develop the qEAM FF are zero 
temperature EOS for different crystal phases of Ta in a wide pressure 
range. We used energy-volume and pressure-volume for bcc Ta from
Section II; fcc-bcc energy difference and fcc pressure for different
volumes from Section II; and first principles A15-bcc energy difference 
obtained by S{\"o}derlind and Moriarty \cite{soderlind98} using full potential
linear muffin-tin orbital method within the GGA approximation with
spin orbit interactions.
Ta is a bcc metal and no pressure-induced phase transition to other 
solid structure has been found experimentally or theoretically. 
Nevertheless, using QM it is possible
to calculate the equation of state for different
crystalline structures although they may not be thermodynamically
stable. Including the EOS of different phases in the FF development
leads to an accurate description of the atomic interactions even 
when the environment of an atom is not that of the stable phase; this
could play a key role to correctly describe defects and non-equilibrium processes.

In Figure 1 we show energy [Figure 1(a)] and pressure [Figure 1(b)] as a 
function of volume for bcc Ta at $T=0$ K. The circles denote LAPW GGA
results and the lines the qEAM FF. Figures 2 and 3 show the same results
for fcc and A15 Ta respectively. In Figure 3(a) open circles show the A15 energy calculated
by S{\"o}derlind and Moriarty, \cite{soderlind98} and the filled circles denote the sum
of the A15-bcc energy difference from Ref. \onlinecite{soderlind98} and our bcc energy from
section II, these are the quantities the FF was actually fitted to. The insets on 
Figures 2(a) and 3(a) show the fcc-bcc and A15-bcc
energy difference as a function of volume obtained with the qEAM FF (lines) and 
from QM (circles). In Figure 4(a) and 4(b) we show energy-volume and pressure-volume
curves for hcp phase respectively; circles denote denote the LAPW GGA results of section 
II and the lines show the qEAM FF results. Note that although the the hcp data was not 
included in the training set for the qEAM FF the agreement is very good. Figures 1 to 4 show 
that the qEAM FF reproduces the zero temperature EOS for the four different phases 
very well.

We also included in the FF training set the {\it ab initio} elastic constants 
from Section II at zero pressure.
Table \ref{table_bcc} shows bcc Ta EOS parameters [zero pressure volume 
($V_0$), bulk modulus ($B_T$), its derivative with respect to pressure 
($B_T^{\prime}$) and the elastic constants] obtained using
the qEAM FF together with the QM values from Section II and the
ones reported in Ref. \onlinecite{soderlind98}. $V_0$, $B_T$, and $B_T^{\prime}$ were obtained
fitting Rose's universal equation of state to the energy-volume data shown in Figure 1. 
The elastic constants $c_s$ and $c_{44}$ were calculated
with the qEAM FF using the tetragonal and orthorhombic strains shown above [equations (1) to (4)].

In Figure 5 we show the elastic constants
[bulk modulus $B_T$, $c_s$ and $c_{44}$]
as a function of pressure obtained with the qEAM FF (filled circles and full lines)
and the LAPW results from Section II. While the agreement in $B_T$ is excellent
and that for $c_s$ is good, qEAM FF greatly underestimates $c_{44}$ for high
pressures; this problem is amplified by the possible electronic phase transition that
leads to a change of behavior of $c_{44}$ at $~\sim$ 150 PGa (see Section II and Ref. \onlinecite{oguz2001}).

In order to estimate the accuracy of the the numerical calculation of elastic constants we 
have calculated $c_{11}$ independently using a uniaxial strain: 
\begin{eqnarray}
\mathbf{a} & = & a ( 1+\epsilon , 0 , 0),\nonumber \\
\mathbf{b} & = & a ( 0, 1 , 0),\nonumber \\
\mathbf{c} & = & a ( 0 , 0 , 1),
\end{eqnarray}
the elastic constant $c_{11}$ is obtained from:
\begin{equation}
E(\epsilon) = E(\epsilon=0) + P V(\epsilon=0) \epsilon + \frac{1}{2}  V(\epsilon=0)  c_{11}  \epsilon^2 + O(\epsilon^3),
\end{equation}
where $E(\epsilon=0)$ is the zero strain energy, $V(\epsilon=0)$ is the volume at zero strain, and P is the 
zero strain hydrostatic pressure.
For zero pressure we calculate $c_{11} = 273.67 GPa$ 
only 0.4 $\%$ higher than the one computed from $c_s$ [equations (1) and (2)] and $B_T$
(calculated form the energy-volume data shown in Figure 1). We find a similar agreement
under compression: for pressure P=109.54 GPa
(corresponding to a volume per atom $V(\epsilon=0) = 13.36 \AA^3$) we obtain $c_{11} = 819.90$ GPa
only 0.8 $\%$ larger than the calculated from $c_s$ and $B_T$.

We have calculated the T$=300$ K EOS using isothermal-isobaric (NPT) MD with
a Hoover \cite{hoover} thermostat and Rahman-Parrinello barostat.
\cite{rahman81} In Table \ref{table_bcc} we show the T=$300$ K zero pressure
volume, bulk modulus and its first derivative
with respect to volume; we also show recent compressibility  data  \cite{cynn}
obtained in a diamond-anvil cell at room temperature and ultrasonic measurements of adiabatic
elastic constants. \cite{katahara}

Zero temperature EOS data obtained with the qEAM FF is available in the supplementary
material. \cite{supp_mat}

\subsection{Vacancy formation energy}

We used experimental values for vacancy formation energy 
and cohesive energy in the training set used to fit the qEAM FF. 
The experimental value for the relaxed vacancy formation energy is $E_{vac} = 2.8$ 
eV. \cite{vac_exp} From this value we estimated the value of the unrelaxed vacancy 
formation energy to be $3$ eV. This unrelaxed value was used to fit the qEAM FF. 
The value for the cohesive energy used is $E_{coh} = 8.10$ eV. \cite{coh_exp}

The volume-relaxed vacancy formation energy is defined in computer 
simulations as:
\begin{equation}
e_{vac}(P)= e_{vac}(N-1,P) -  \frac{N-1}{N} e_{xtal}(N,P),
\label{vacancyenergy}
\end{equation}
where $e_{xtal}(N,P)$ is the energy of the perfect crystal with N atoms
at pressure $P$ and $e_{vac}(N-1,P)$ is the energy corresponding to the
N-1 atoms system with a vacancy at pressure $P$ where the atomic positions
are not allowed to relax. The relaxed vacancy formation energy is defined
in the same way but with $e_{vac}(N-1,P)$ being the atom-relaxed energy
of the system with a vacancy.

The volume-relaxed vacancy formation energy obtained using the qEAM
FF is 3.10 eV, in very good agreement with the target value (3 eV) and only slightly
lower than the {\it ab initio} value $3.25$eV. The relaxed vacancy formation
energy obtained with the qEAM FF is $2.95$ eV. Previous work by Korhonen,
Puska and Nieminen, using a DFT full potential linear muffin-tin orbital method with LDA 
gives $3.49$ eV for
the unrelaxed vacancy formation energy; \cite{korhonen95} Satta, Willaime and Gironcoli
\cite{satta99} (using DFT plane Waves LDA) calculated $3.51$ eV and $2.99$ eV for
unrelaxed and relaxed vacancy formation energies; recently S\"oderlind,
Yang, Moriarty and Wills (also using the full potential linear muffin-tin orbital method
with GGA) obtained $3.74$ eV for the
volume relaxed vacancy formation energy and $3.2$ for the relaxed vacancy
formation energy. Table \ref{table_vac} summarizes our vacancy formation
energies together with previous theoretical results and experimental
data.

In Figure 6(a) we show the volume-relaxed vacancy formation energy ($e_{vac}$)
as a function of pressure (P); the thick solid line shows qEAM results and 
circles denote QM results of Section II. We used a simulation cell containing
54 atoms for the perfect crystal case with periodic boundary conditions.
Vacancy formation enthalpy is defined in the same way as $e_{vac}(P)$ [eq. (\ref{vacancyenergy})]
replacing energy with enthalpy.
In Figure 6(b) we plot the vacancy formation enthalpy $h_{vac}$ with respect 
to pressure.
The difference between the vacancy formation enthalpy as obtained by 
qEAM FF and QM, is due to the difference in the vacancy formation volume obtained 
from the two methods. The vacancy formation volume [$v_{vac}(P)$] is, again, defined 
in the same way as $e_{vac}(P)$ [eq. (\ref{vacancyenergy})] replacing energy with volume. 
In Figure 7 we plot the vacancy formation volume 
($\Omega{^{f}}_{vac}$), with respect to pressure. While the pressure dependence of
$v_{vac}(P)$ calculated using the qEAM FF is very similar to the MB-PS calculation, our
Force Field overestimates the vacancy formation volume resulting in higher vacancy
formation enthalpy for compressed states.

In order to calculate the vacancy formation enthalpy at finite temperatures
as a function of pressure we performed NPT MD simulations using a
cell containing $N=1458$ atoms with periodic boundary conditions.
We performed simulations at 9 different volumes (19, 18.36, 18, 17, 16, 15,
14, 13, 11 $\AA^3$); for each volume we started with the relaxed structure
a $T=0$ K and heat the system in $100$ K steps; for each temperature we
performed $25$ ps MD simulations and used the last $20$ ps for measurements.
In Figure 8 we show vacancy formation enthalpy as a function of pressure for
$T=0 K$ (full atomic relaxation), $T=300$ K, the volume-relaxed enthalpy is also
shown for comparison; the zero pressure values are also shown in Table \ref{table_vac}.
The fundamental data used to calculate vacancy energy, enthalpy and volume
can be found in the supplementary material. \cite{supp_mat}

A very important quantity, which determines vacancy mobility, is the vacancy
migration energy barrier ($E_{vac}^{mig}$). We calculate $E_{vac}^{mig}$ using the 
qEAM FF by marching
a nearest neighbor atom towards the position of the vacancy in short steps. 
At each step the position of the marching atom is fixed, as well as that of 
a distant, reference, atom, and the positions of all the other atoms are relaxed
at constant pressure. In this way we obtain the optimum migration path and
energy as a function of displacement.
In Figure 9 we show the energy as a function the position of the marching
atom for zero pressure and zero temperature; we obtain a vacancy migration 
energy $E_{vac}^{mig} \sim 1.093$ eV. 
The activation energy for self diffusion is defined as 
$Q = E_{vac}^{f} + E_{vac}^{mig}$, using the qEAM FF we obtain $Q = 4.028$ eV in 
good agreement with the experimental value $3.8 \pm 0.3$ eV, \cite{vac_mig}
and {\it ab initio} calculations \cite{satta99} which give $3.82$ eV,
see Table \ref{table_vac}. At T=300 K we obtain a vacancy migration energy
of 1.1 $\pm$ 0.5 eV, see Table \ref{table_vac}, very similar to the zero temperature
value.

\subsection{Energetics of homogeneously sheared bcc crystal}

Zero temperature, first principles energetics of a homogeneously sheared 
bcc crystal in the observed twinning mode was also included in the qEAM 
training set.

The ideal shear strength is defined to be the stress separating elastic
and plastic deformation when a homogeneous shear is applied to a perfect
crystal. It gives an upper bound for the shear strength of the material.
The shear transformation is in the direction of the observed twining mode
and deforms the crystal into itself. \cite{soderlind98,paxton91}
For bcc crystal we use the following transformation of the cell vectors:
\cite{soderlind98,paxton91}
\begin{eqnarray}
{\mathbf a} = \frac{1}{2} [\bar1 1 1 ] + \frac{s}{\sqrt{18}} [ \bar1 \bar1 1], \nonumber\\
{\mathbf b} = \frac{1}{2} [1 \bar 1 1 ] + \frac{s}{\sqrt{18}} [ \bar1 \bar1 1], \nonumber\\
{\mathbf c} = \frac{1}{2} [1 1 \bar1 ],
\end{eqnarray}
when the shear variable $s$ is equal to the twinning shear $s = s_{tw} = 2^{-1/2}$ the lattice 
vectors [$a = 1/3 [\bar 2 1 2]$, $b=1/3 [1 \bar 2 2]$ and 
$c =\frac{1}{2} [1 1 \bar1 ]$] form a bcc structure, twin of the initial one.

In this way one can calculate the energy along the shear path, 
\begin{equation}
W(s) = e(V,s) - e(V,s=0),
\end{equation}
where $e(V, s)$ is the energy per atom of the deformed system and 
$e(V,s=0)$ is the perfect crystal energy. The energy barrier associated
with this transformation is $W_{max} = W(s = 0.5)$.
The corresponding stress is defined as:
\begin{equation}
\tau(s) = \frac{1}{V} \frac{dW(s)}{d s}.
\end{equation}
The ideal shear strength ($\tau_{max}$) is defined as the maximum stress along the path.
In Figure 10 we show energy and stress as a function of shear using the qEAM FF for
zero pressure volume, see also Ref. \onlinecite{supp_mat}.

S{\"o}derlind and Moriarty \cite{soderlind98} calculated $W(s)$ and
$\tau(s)$ for Ta at different volumes, from first principles.
In developing the qEAM FF we used $W_{max}$ for V=$17.6186 \AA^3$ and V=$10.909 \AA^3$ as
part of the training set. In Table \ref{table_shear} we show a comparison between the first
principles results \cite{soderlind98} and the ones obtained using the qEAM FF.
We can see that the qEAM results are  is in very good agreement with the {\it ab initio} calculations.

\subsection{Surface energy}

The unrelaxed (100) surface energy using the qEAM FF is $1.971$ J/m$^2$,
lower than the first principles values of $3.097$ J/m$^2$ of Ref. \onlinecite{surf_qm} and
3.23 J/m of Ref. \onlinecite{mailhiot95}. Low surface
energy is a common problem in EAM-like force fields, see Ref. \onlinecite{guellil92}.
The zero temperature experimental estimate of the surface energy (averaged over 
different surfaces) is 2.902 J/m$^2$. \cite{tyson77}

\section{Thermal expansion}

Thermal expansivity is an important materials property that can be
calculated directly from MD simulations. 
In order to calculate the lattice constant as
a function of temperature for zero pressure we performed NPT MD 
simulations with a computational cell containing 1024 atoms increasing
the temperature by 100 K every 25 ps at zero applied pressure.
For each temperature the first 5 ps were taken as thermalization
and the remaining 20 ps were used for measurements. \cite{supp_mat}
{\it ab initio} MD simulations are very time consuming and only feasible for small
systems and short times; thus in order to calculate the thermal expansion
from first principles we take the following approach. \cite{cohen_thermal}
The Helmholtz free energy can be written as:
\begin{equation}
F(V,T) = E_0(V) + F_{el}(V,T) + F_{vib},
\end{equation}
where $E_0(V)$ is the zero temperature energy (Section II), $F_{el}(V,T)$ is
the electronic contribution and $F_{vib}$ is the vibrational part of the
free energy. The electronic contribution is obtained using quantum statistical
mechanics. The vibrational part is obtained using the particle in a cell method,
\cite{pic} further detail of our calculations can be found in Ref. 
\onlinecite{cohen_thermal}. This calculations were
performed using the mixed basis pseudopotential method and a cell containing
54 atoms.

In Figure 11(a) we show the thermal expansivity 
($\alpha(T) = 1/V \partial V / \partial T$) as a function of temperature
obtained form our MD simulations (circles), first principles results (dashed-dotted line), 
as well as the experimental values (full and dashed lines). \cite{touloukian} 
Figure 11(b) shows the linear thermal expansion 
$(a - a_0)/a_0$ (where $a_0$ is the T$=300$ K lattice constant) as a function
of temperature obtained using the qEAM FF (circles) as well as the experimental
values. \cite{touloukian} The high temperature experimental results (shown as
dashed lines) represent provisional experimental data.\cite{touloukian}
Both MD and the particle in cell methods are based on classical mechanics so none of the
them are expected to capture the low temperature behavior where the differences between
quantum and classical statistical mechanics are important. The force field 
results agree with experimental data well; for example the qEAM FF underestimates the change
in lattice parameter from $T=300$ K to $T=2000$ K by less than $0.2\%$. This is an important result since the 
thermal expansion in related to anharmonicities of the internal energy. Our {\it ab initio}
data also agree well with the experimental results; we slightly overestimate thermal 
expansivity for temperatures lower than T=2000 K.

\section{Melting curve of T\lowercase{a}}

We studied melting of Ta using one phase and two-phase MD simulations
with the qEAM force field. 

In Figure 12 we show the results of zero pressure heating a solid sample 
until it melts and then cooling the melt. The system contains 1024 atoms with
periodic boundary conditions with a heating and cooling rate of 100 K per 
25 ps. In Figure 12 we plot enthalpy [Figure 12(a)] and volume [Figure 12(b)] 
as a function of temperature. The solid superheats on heating and the
liquid undercools as expected for a small, periodic system with
no defects to act as nucleation centers and high heating and cooling rates.

In order to overcome some of these problems we calculate the
melting temperature using the ``two phase technique''.
We place pre-equilibrated liquid and pre-equilibrated solid samples in a single 
computational cell. Once
this initial configuration is built we perform TPN MD (using a
Hoover thermostat and Rahman-Parinello barostat)
simulation and observe which phase grows. If the simulation temperature T 
is lower than the melting temperature at the given pressure [$T_{melt}(P)$] 
the liquid will start crystallizing,
on the other hand the solid will melt if $T > T_{melt}(P)$.

Given that the liquid-solid phase transition is first order, i.e. the
enthalpy of the solid and liquid in equilibrium differ by a finite
amount, it is very easy to know whether the system is crystallizing
or melting by analyzing the time evolution of the total potential
energy during the MD run. In Figure 13 we show potential energy
as a function of time for zero pressure two-phase simulations
at different temperatures. We can see that at T=3100 K (dotted line in
Figure 13) the potential energy decreases with time; this means that the solid
phase is growing and $T_{melt}> 3100$ K. On the other hand,
for $T = 3200$ K the energy grows, the system is, then, melting and
$T_{melt} < 3200$ K. At T = 3150 K the energy is rather constant
and both phases are close to equilibrium. This value for the zero
pressure melting is in very good agreement with experimental results
which range from $3213$ K to $3273$ K. \cite{hultgren}
This is a very important validation of the FF, taking into account that 
only zero temperature data was used in its development. 

The slope of the melting curve is given by the Clausius Clapeyron equation:
\begin{equation}
\left[ \frac{dP}{dT} \right]_{melt} = \frac{1}{T} \frac{\Delta H}{\Delta V},
\end{equation}
where $\Delta H$ and $\Delta V$ are the enthalpy and volume difference 
between the liquid and solid in equilibrium respectively.
From our MD runs, see Figure 12, the slope of the melting curve at
zero pressure is $dT/dP =  92.8$ K/GPa, larger than the the experimental value 
$dT/dP = 60 \pm 10$ K/GPa.\cite{shaner77}

Using the two phase simulation procedure we calculated the melting temperature
for various pressures up to $300$ GPa. In Figure 14 we show the
melting curve for Ta, see also Ref. \onlinecite{supp_mat}. To the best of our 
knowledge this is the first 
calculation of melting temperature in Ta for such a wide pressure range.
Zero pressure experimental values \cite{hultgren} are also shown in Figure 14 as empty diamonds.
High pressure melting of Ta has been studied experimentally via shock compression; \cite{brown83}
melting is identified as a change in the velocity of the rarefication wave
(from the longitudinal to the bulk sound velocity). The transition was found to occur in the
pressure range from $\sim$250 GPa to 295 GPa.\cite{brown83} The calculation of the temperature
in shock experiments (i.e. along the Hugoniot) is difficult; the electronic contribution to the 
specific heat has a very strong effect on the melting temperature. \cite{brown83}
Simple models for the electronic behavior lead to very different temperatures: using the free 
electron gas model the melting temperature is $\sim$10000 K while considering
band electrons give $T_{melt}\sim$ 7500 K, see Ref.\onlinecite{brown83}. These experimental values 
are shown in Figure 14 are empty circles. Using the accurate {\it ab initio} thermal equation of 
state obtained using the
methods described above and the Rankine-Hugoniot equation Cohen and G{\"u}lseren calculated pressure-volume
and temperature-pressure curves for Ta under shock conditions.\cite{cohen_thermal} 
This calculation leads to a melting temperature
(using the experimental melting pressure P$_{melt}$=375 GPa) of 8150 K (see square in Figure 14). 
Our MD results are in good agreement with the high pressure calculation of the melting temperature based on
shock experiments and {\it ab initio} calculations.

\section{Conclusions}

We have presented here a general strategy to derive accurate classical force fields
based on {\it ab initio} QM mechanical calculations for metallic systems.
Force fields allow calculations on systems containing millions of atoms, providing a 
means to study from an atomistic point of view a variety of processes relevant
to the mechanical and thermodynamical properties of metals, such as phase transitions, dislocations
dynamics and interactions, failure, crack propagation, etc. 

We showed that the qEAM FF describes with good accuracy the EOS of Ta for different 
crystal phases [bcc (coordination number 8), fcc (coordination number 12), and A15
(mixed coordination numbers)], and also elastic constants, vacancy formation 
energy, and energetics of the deformed bcc lattice in the twinning direction.
A critical point in our approach is that in developing the force field we use not only 
the thermodynamically stable phase but also high energy phases and large strains; a
force field that can correctly describe such structures should be appropriate for
simulations of defects and non-equilibrium processes.

The large amount of QM data used to derive our force field gives an important measure of 
its quality. 

We used the qEAM FF with MD to calculate the melting curve for Ta in a wide pressure range.
The zero pressure melting temperature obtained from our simulations
$T_{melt} = 3150 \pm 50$ K is in very good agreement with the experimental
result of $3290 \pm 100$ K. This is an important validation of our approach given the fact that the
qEAM FF was derived using zero temperature data.

First-priciples-based force fields represent an important step in {\it ab initio} multiscale modeling 
of materials.
We have used the qEAM FF to study spall failure, \cite{strachan2000} crack propagation, and dislocation 
properties such as core structure and energy, Peierls stress, and kink formation energies
\cite{wang2000} with systems containing as many as 50,000
atoms. We used such calculations with a micromechanical model of plasticity developed by
Stainer, Cuiti\~no and Ortiz \cite{tamodel} to develop a multiscale model of single crystal plasticity in
Ta.\cite{multiscale} We are currently performing shock simulations in systems containing more than
a million atoms using a parallel MD code; for a system containing 1,098,500 atoms
1 ps of MD simulations (1000 steps) takes approximately one hour on an SGI Origin 2000 
computer using 128 250-MHz R10000 processors.

\acknowledgements

This research was funded by a grant from DOE-ASCI-ASAP (DOE W-7405-ENG-48).  
The facilities of the MSC are also supported
by grants from NSF (MRI CHE 99), ARO (MURI), ARO (DURIP), NASA, Kellogg,
Dow Chemical, Seiko Epson, Avery Dennison, Chevron Corp.,
Asahi Chemical, 3M, GM, and Beckman Institute. Thanks to D. Singh and
H.  Krakauer for use of their LAPW code. We also thank Per S{\"o}derlind and
John Moriarty for providing us with LMTO data. Computations were partially
performed on the Cray SV1 at the Geophysical Laboratory, supported by NSF
grant EAR-9975753 and the W.\ M.\ Keck Foundation.

\newpage

\begin{table}
\caption{EOS parameters for bcc Tantalum.}
\label{table_bcc}
\begin{tabular}{l|c|c|c|c|c|c}
&$V_0(A^3)$ & $B_T$(GPa) & $B_T^{\prime}$ & $c_{11}$ (GPa) & $c_{12}$ (GPa) &$c_{44}$ (GPa) \\
\tableline
&\multicolumn{6}{c}{{\bf Theory} (0 K)}\\
\tableline
This work (LAPW-GGA)   &  18.33  & 188.27 & 4.08  &  245.18  &  159.8  &  67.58 \\
This work (qEAM FF)    &  18.36  & 183.04    & 4.16  &  272.54     &  137.57    &  69.63 \\
FP LMTO GGA SC \tablenotemark[1]   &  17.68  & 203    &  -    &  281     &  163    &  93    \\
\tableline
&\multicolumn{6}{c}{{\bf Theory} (300 K)}\\
\tableline
This work (qEAM FF)    &  18.4   & 176    & 4.9   &    -     &   -     &  -  \\
\tableline
&\multicolumn{6}{c}{{\bf Experiment} (300 K)}\\
\tableline
Diamond Anvil Cell \tablenotemark[2] &  18.04   & 194.7$\pm$4.8  &  3.4 &   -      &    -    &   -  \\
Ultrasonic  \tablenotemark[3] &   -      &   -            &   -  &  266     &  160.94 &  82.47 \\
\end{tabular}
\tablenotetext[1]{Full potential linear muffin-tin orbital calculations, P. S{\"o}derlind and J. A. Moriarty. \cite{soderlind98}}
\tablenotetext[2]{H. Cynn and C. Yoo. \cite{cynn}}
\tablenotetext[3]{Adiabatic elastic constants at 25$^\circ$ C, Katahara, Manghnani, and Fisher. \cite{katahara}}
\end{table}

\begin{table}
\caption{qEAM paramters for Ta: two body term. The units of $b_i$ are eV$\AA^{-(4+i)}$.}
\label{table_ff1}
\begin{tabular}{|c|c|c|c|c|}
 $r_m~\AA $     &    $b_0$   &     $b_1$    &    $b_2$   &    $b_3$    \\
\tableline
4.81253968 & 6.50281587 & -11.26455130 & 8.01451544 & -2.97299223 \\
\tableline
\tableline
 $b_4$     &    $b_5$   &     $b_6$    &    $b_7$   &      -      \\
\tableline
0.60004206 &-0.06222106 &   0.00258801 & -0.00000504&      -      \\
\end{tabular}
\end{table}

\begin{table}
\caption{qEAM paramters for Ta: embedding energy}
\label{table_ff2}
\begin{tabular}{|c|c|c|c|c|}
   $a_1$     &   $ a_2$   & $\alpha$ (1/\AA) &  $\beta$   &  $a_0$ (\AA)   \\
\tableline
0.07293238   & 0.15781672 &    21.79609053    & 7.79329426 & 3.32389219   \\
\tableline
\tableline
$E_{coh}$(eV)&$B_T$ (GPa)&     $\lambda$     &   $f_3$    &   $f_4$      \\
\tableline
8.15420437   & 183.035404 &     0.20782789    & -0.00717801&  -0.00000504 \\
\end{tabular}
\end{table}

\begin{table}
\caption{Volume-relaxed ($e_{vac}^{vr}$), full-relaxed Vacancy formation ($e_{vac}^{ar}$), vacancy 
migration energies ($e_{vac}^{mig}$), and activation energy for self diffusion (Q).}
\label{table_vac}
\begin{tabular}{|l|c|c|c|c|}
                                 &$e_{vac}^{vr}$ (eV)&$e_{vac}^{ar}$(eV)&$e_{vac}^{mig}$(eV)&  Q (eV)\\
\tableline
&\multicolumn{4}{c|}{{\bf Theory} (0 K)}\\
\tableline
This work (qEAM FF)               &  3.10   &   2.935  & 1.093   &  4.028 \\
This work (MB-PS)                 &  3.25   &     -    &   -    &   -    \\
FP LMTO GGA SC \tablenotemark[1]  &  3.74   &   2.20   &   -    &   -    \\
Plane waves LDA \tablenotemark[2] &  3.51   &   2.99   & 0.83   &  3.82  \\
FP LMTO LDA    \tablenotemark[3]  &  3.49   &     -    &   -    &    -   \\
\tableline
&\multicolumn{4}{c|}{{\bf Theory and Experiment} (300 K)}\\
\tableline
This work (qEAM FF)              &   -   &     3.0 $\pm$ 0.05    &  1.1 $\pm$ 0.05  &  4.1 $\pm$ 0.1 \\
Experiment \tablenotemark[4]     &   -   &     2.8 $\pm$ 0.6     &   -              &  3.8 $\pm$ 0.3 \\
\end{tabular}
\tablenotetext[1]{P. S{\"o}derlind and J. A. Moriarty.\cite{soderlind00}}
\tablenotetext[2]{Satta, Willaime, and Gironcoli.\cite{satta99}}
\tablenotetext[3]{Unrelaxed value by Korhonen, Puska, and Nieminer.\cite{korhonen95}}
\tablenotetext[4]{Ref. \onlinecite{vac_exp} for vacancy energy and Ref. \onlinecite{vac_mig} for activation energy}
\end{table}

\begin{table}
\caption{Shear deformation in the observed twinning mode in Ta.}
\label{table_shear}
\begin{tabular}{|c|c|c|c|c|}
 Volume (\AA$^3$)  &  $W_{max}$ (eV) & $\tau_{max}$ & $W_{max}$ (eV) & $\tau_{max}$ \\
\tableline
  & \multicolumn{2}{c|}{This work (qEAM FF)} & \multicolumn{2}{c|}{FP LMTO GGA SC \tablenotemark[1]} \\
\tableline
18.36        & 0.188  & 7.14   &   -     &  -    \\
17.618602    & 0.2    & 8.0    & 0.194   & 7.37  \\
15.143996    & 0.26   & 12.05  & 0.276   & 12.4  \\
10.9090116   & 0.43   & 28.2   & 0.566   & 36.2  \\
\end{tabular}
\tablenotetext[1]{P. S{\"o}derlind and J. A. Moriarty. \cite{soderlind98}}
\end{table}

\newpage
{\bf FIGURE CAPTIONS}

Figure 1. Zero temperature EOS for bcc Ta, LAPW GGA and qEAM FF results. Energy 
[Figure 1(a)] and pressure [Figure 1(b)] as a function of volume. 
Circles denote LAPW GGA results and lines show qEAM FF results.

Figure 2. Zero temperature EOS for fcc Ta, LAPW GGA and qEAM FF results. Energy 
[Figure 2(a)] and pressure [Figure 2(b)] as a 
function of volume. Circles denote LAPW GGA results and
lines show qEAM FF results. The inset of Figure 2(a) shows fcc-bcc
energy difference as a function of volume (circles joined by dots 
denote LAPW GGA results and line qEAM FF).

Figure 3. Zero temperature EOS for A15 Ta, {\it ab initio} and qEAM FF results.  Energy 
[Figure 3(a)] and pressure [Figure 3(b)] as a 
function of volume. Open circles with dotted line denote full potential muffin-tin
orbital calculations results Ref. 
\onlinecite{soderlind98}; filled circles show the sum of A15-bcc energy
difference from \cite{soderlind98} and our calculation of bcc energy 
using LAPW GGA method (section II). The full line shows qEAM FF results.
The inset of Figure 3(a) shows A15-bcc energy difference as a function 
of volume (circles with dotted line denote QM results \cite{soderlind98} and 
the line qEAM FF).

Figure 4. Zero temperature EOS for hcp Ta, LAPW GGA and qEAM FF results. Energy 
[Figure 4(a)] and pressure [Figure 4(b)] as a 
function of volume. Circles denote LAPW GGA results, section II, and
lines show qEAM FF results. The inset of Figure 4(a) shows hcp-bcc
energy difference as a function of volume (circles with dotted line denote 
LAPW GGA results and line qEAM FF).

Figure 5. Zero temperature elastic constants for Ta, LAPW GGA and qEAM FF results. 
Circles show bulk modulus [$(c_{11} + 2 c_{12})/3$]; diamonds show $c_{44}$ and
squares represent $c_s = (c_{11}-c_{12})/2$. qEAM FF results are
shown with filled symbols and full lines and {\it ab initio} LAPW results with 
open symbols and dashed lines.

Figure 6. Volume relaxed vacancy formation energy (a) and enthalpy (b)
as function of pressure. Dashed lines represent {\it ab initio} MB-PS results (section
II) and full lines show qEAM results.

Figure 7. Vacancy formation volume as function of pressure. Dashed 
lines represent {\it ab initio} MB-PS results (section II) and full lines show qEAM results.

Figure 8. Vacancy formation enthalpy as function of pressure using the qEAM FF. 
Solid line shows the volume relaxed result; the dotted line the T=0 K fully
relaxed results; dashed line is T=$300$ K result.

Figure 9. Vacancy migration energy using qEAM FF. Energy as a function of
position of the marching atom at T=0 K and zero pressure. The vacancy migration 
energy is 1.093 eV.

Figure 10. Ideal shear strength of Ta using qEAM FF at zero temperature and volume
V=18.36 $\AA^3$. We show energy $W(s)$ [Figure 10 (a)] and stress $\tau(s)$ 
[Figure 10 (b)] as a function of shear.

Figure 11. Thermal expansion in Ta. (a) thermal expansivity as a function of temperature; 
circles represent qEAM FF results, the dashed-dotted lines shows mixed basis pseudopotantial
calculations using the particle in a cell method, and the line denotes experimental results 
from Ref. \onlinecite{touloukian}. 
(b) Linear thermal expansion of Ta. $(a - a_0)/a_0$ as a function
of temperature; circles represents qEAM FF results and line denote experimental
results from Ref. \onlinecite{touloukian}. The high temperature experimental data (dashed lines) are
provisional values.\cite{touloukian}

Figure 12. Melting of tantalum using the qEAM FF. Enthalpy (a) and volume (b) as a function of 
temperature for zero pressure; heating of bcc Ta (lower branches) and 
cooling of liquid Ta (higher branches). Heating and cooling rates are 100 K
per 25 ps.

Figure 13. Ta melting using the qEAM FF. Two phase simulations. Time evolution of the potential
energy in TPN MD at zero pressure for different temperatures.
For T=3100 K (dotted line) the potential energy decreases with time,
i.e. the system is crystallizing; for T=3200 K (dashed line) the potential energy
grows denoting melting. For $T=3150 \sim T_{melt}(P=0)$ the
potential energy remains constant; in this case the solid and liquid phases coexist
in equilibrium.

Figure 14. Ta Melting qEAM FF. Melting curve for Ta up to P$=300$ GPa obtained using the
two-phase simulation technique. We also show the experimental zero pressure melting temperature
\cite{hultgren} (circles) and the results of shock melting. \cite{brown83}


\begin{references}

\bibitem{soderlind98} 
Per S{\"o}derlind and John A. Moriarty,
{\sl Phys. Rev. B} {\bf 57}, 10340 (1998).

\bibitem{soderlind00} Per S{\"o}derlind, L. H. Yang, John A. Moriarty, and J. M. Wills,
{\sl Phys. Rev. B} {\bf 61}, 2579 (2000).

\bibitem{satta99} Alessandra Satta, F. Willaime, and Stefano de Gironcoli,
{\sl Phys. Rev. B} {\bf 60}, 7001 (1999).

\bibitem{sohrab00} Sohrab Ismail-Beigi and T. A. Arias,
{\sl Phys. Rev. Lett.} {\bf 84}, 1499 (2000).

\bibitem{mori_momelt} John A. Moriarty,
{\sl Phys. Rev. B} {\bf 49}, 12431 (1994).

\bibitem{mori_modisl} W. Xu and John A. Moriarty,
{\sl Phys. Rev. B} {\bf 54}, 6941 (1996).

\bibitem{mori_ta} John A. Moriarty, W. Xu, P. S{\"o}derlind, J. Belak, L. H. Yang and J. Zhu,
{\sl J. Eng. Mater. Tech.} {\bf 121}, 120 (1999). 

\bibitem{daw84} M. S. Daw and M. I. Baskes,
{\sl Phys. Rev. Lett.} {\bf 50}, 1285 (1983);
{\sl Phys. Rev. B} {\bf 29}, 6443 (1984).

\bibitem{chanta96} Somchart Chantasiriwan and Frederick Milstein,
{\sl Phys. Rev. B} {\bf 53}, 14080 (1996).

\bibitem{johnson89} R. A. Johnson and D. J. Oh,
{\sl J. Mater. Res.} {\bf 4}, 1195 (1989)

\bibitem{guellil92} A. M. Guellil and J. B. Adams,
{\sl J. Mater. Res.} {\bf 7}, 639 (1992).

\bibitem{MEAM} M. I. Baskes,
{\sl Phys. Rev. Lett.} {\bf 59}, 2666 (1987);
M. I. Baskes, J. S. Nelson, and A. F. Wright,
{\sl Phys. Rev. B} {\bf 40}, 6085 (1989).

\bibitem{MEAM_metal} M. I. Baskes,
{\sl Phys. Rev. B} {\bf 46}, 2727 (1992);
Byeong-Joo Lee, M. I. Baskes, Hanchul Kim, and Yang Koo Cho,
{\sl Phys. Rev. B} {\bf 64}, 184102 (2001).

\bibitem{weikrakauer}
S.H. Wei and H. Krakauer,
{\sl Phys. Rev. Lett.} {\bf 55},  1200-1203  (1985).

\bibitem{singh} 
D.J. Singh,
{\it Planewaves, Pseudopotentials, and the LAPW Method}
(Kluwer Academic Publishers, Boston, 1994).

\bibitem{PBE} 
John P. Perdew, Kieron Burke, and Matthias Ernzerhof
{\sl Phys. Rev. Lett.} {\bf 77}, 3865 (1996).

\bibitem{monpack} 
H.J. Monkhorst and J.D. Pack,
{\sl Phys. Rev. B} {\bf 13}, 5188 (1976).

\bibitem{supp_mat} See our web site (http://www.wag.caltech.edu/publications/papers/).
In addition see EPAPS for supplementary tables.

\bibitem{vinet} 
P. Vinet, J. Ferrante, J.R. Smith and J.H. Rose,
{\sl J. Phys. C: Solid State Phys.} {\bf 19}, L467 (1986);
P. Vinet, J.H. Rose, J. Ferrante and J.R. Smith,
{\sl J. Phys.: Condens. Matter} {\bf 1}, 1941 (1987).

\bibitem{cynn} Hyunchae Cynn and Choong-Shik Yoo,
{\sl Phys. Rev. B} {\bf 59}, 8526 (1999).

\bibitem{oguz2001}
O. G{\"u}lseren and R. E. Cohen,
{\sl Phys. Rev. B} (to appear).

\bibitem{katahara} 
K.W. Katahara, M.H. Manghnani and E.S. Fisher,
{\sl J.Phys. F: Metal Phys.} {\bf 9}, 773 (1979);
K.W. Katahara, M.H. Manghnani and E.S. Fisher,
{\sl J. Appl. Phys.} {\bf 47}, 434 (1976).

\bibitem{cohenta}
R.E. Cohen, O. G\"{u}lseren and R.J. Hemley,
Amer. Mineral. {\bf 85},  338-344 (2000).

\bibitem{sonali}
S. Mukherjee, O. G\"{u}lseren, and R.E. Cohen, unpublished.

\bibitem{oguz}  
O. G\"{u}lseren, D.M. Bird and S.E. Humphreys,
{\sl Surface Science} {\bf 402--404}, 827 (1998).

\bibitem{pw91} 
J.P. Perdew and Y. Wang,
{\sl Phys. Rev. B} {\bf 46}, 6671 (1992).

\bibitem{daw93} Murray S. Daw, Stephen M. Foiles, and Michael I. Baskes,
{\sl Mat. Sci. Rep.} {\bf 9}, 251 (1993). 

\bibitem{surf_qm} L. Vitos, A. V. Ruban, H. L. Skriver, and J. Koll'ar,
{\sl Surf. Sci.} {\bf 411}, 186 (1998).

\bibitem{hoover} W. G. Hoover,
{\sl Phys. Rev. A} {\bf 31}, 1695 (1985).

\bibitem{rahman81} M. Parinello and A. Rahman,
{\sl J. Appl. Phys.} {\bf 52}, 7182 (1981).

\bibitem{vac_exp} K. Maier, M. Peo, B. Saile, H. E. Schaefer, and A. Seeger,
{\sl Philos. Mag. A} {\bf 40}, 701 (1979)

\bibitem{coh_exp} F.R. de Boer et at.,
{\it Cohesion in metals: transition metal alloys}
(Elsevier Scientific Pub., New York, 1988).

\bibitem{korhonen95} T. Korhonen, M. J. Puska, and R. M. Nieminen,
{\sl Phys. Rev. B} {\bf 51}, 9526 (1995).

\bibitem{vac_mig} D. Weiler, K. Maier, and H. Mehrer, in
{\it Diffusion in Metals and Alloys}, edited by 
F. J. Kedves and D. L. Beke, (Trans Tech Publications, Aedermannsdorf, 1983).

\bibitem{paxton91} A. T. Paxton, P. Gumbsch, and M. Methfessel,
{\sl Philos. Mag. Lett.}  {\bf 63}, 267 (1991).

\bibitem{mailhiot95} Christine Wu, L. H. Yang, John E. Klepeis, and C. Mailhiot,
{\sl Phys. Rev. B}, {\bf 52}, 11784 (1995).

\bibitem{tyson77} W. R. Tyson and W. A. Miller,
{\sl Surf. Sci.} {\bf 62}, 267 (1977).


\bibitem{cohen_thermal} R.E. Cohen and O. Gulseren,
{\sl Phys. Rev. B}, {\bf 63}, 224101 (2001).

\bibitem{pic}
A.C. Holt, W.G. Hoover, S.G. Gray and D.R. Shortle, {\sl Physica} {\bf 49},
61 (1970); F.H. Ree and A.C. Holt, {\sl Phys. Rev. B} {\bf 8}, 826 (1973);
K. Westera and E.R. Cowley, {\sl Phys. Rev. B} {\bf 11}, 4008 (1975); E.R.
Cowley, J. Gross, Z. Gong and G.K. Horton, {\sl Phys. Rev. B} {\bf 42}, 3135
(1990).

\bibitem{touloukian} Y. S. Touloukian, R. K. Kirby, R. E. Taylor, and P. D. Desai,
{\it Thermophysical Properties of Matter (Thermal Expansion-Metallic 
Elements and Alloys, Vol. 12)} (Plenum Press, New York, 1975).

\bibitem{hultgren} R. Hultgren, P. D. Desai, D. T. Hawkins, M. Gleiser, 
K. K. Kelley, and D. D. Wagman,
{\it Selected values of the thermodynamical properties of the elements}
(American Society for Metals, Metals Park, Ohio, 1973).

\bibitem{shaner77} John W. Shaner, G. Roger Gathers, Camille Minichino,
{\sl High Temp. High Press}, {\bf 9}, 331 (1977).

\bibitem{brown83} J. M. Brown and J. W. Shaner, in
{\it Shock Waves in Condensed Matter-1983}, edited by J. R. Asay,
R. A. Graham, and G. K. Straub, (Elsevier Science Pub. New York, 1984).

\bibitem{strachan2000} Alejandro Strachan, Tahir \ccc, William A. Goddard, III,
{\sl Phys. Rev. B} {\bf 63} 060103(R) (2001).

\bibitem{wang2000} Guofeng Wang, Alejandro Strachan, Tahir \ccc, William A. Goddard, III,
{\sl Mat. Sci. Eng. A} {\bf 309}, 133 (2001).

\bibitem{tamodel} A.~M Cuiti\~no L.~Stainer and M.~Ortiz,
{\sl J. Mech. and Phys. Solids} (to be published).

\bibitem{multiscale} A. M. Cuiti\~no, L. Stainier, Guofeng Wang, Alejandro Strachan,
Tahir \ccc, William A. Goddard, III, and M. Ortiz, {\sl J. Comp. Aided Mat. Design} (to be published).

\end{references}
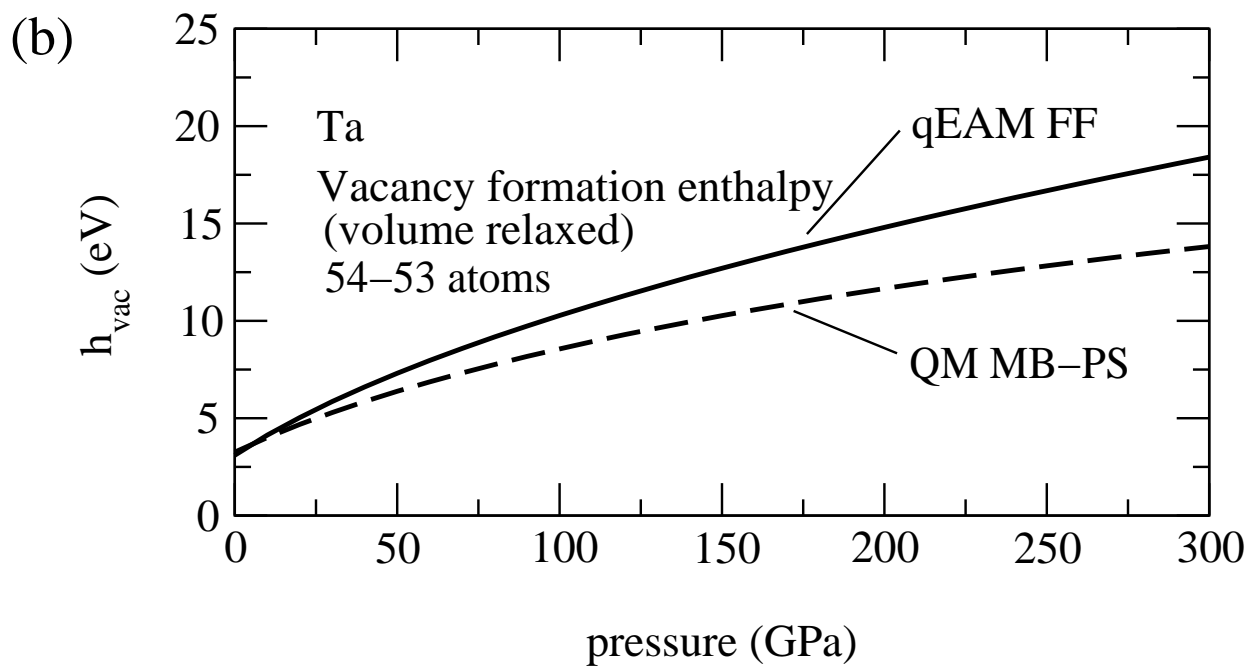
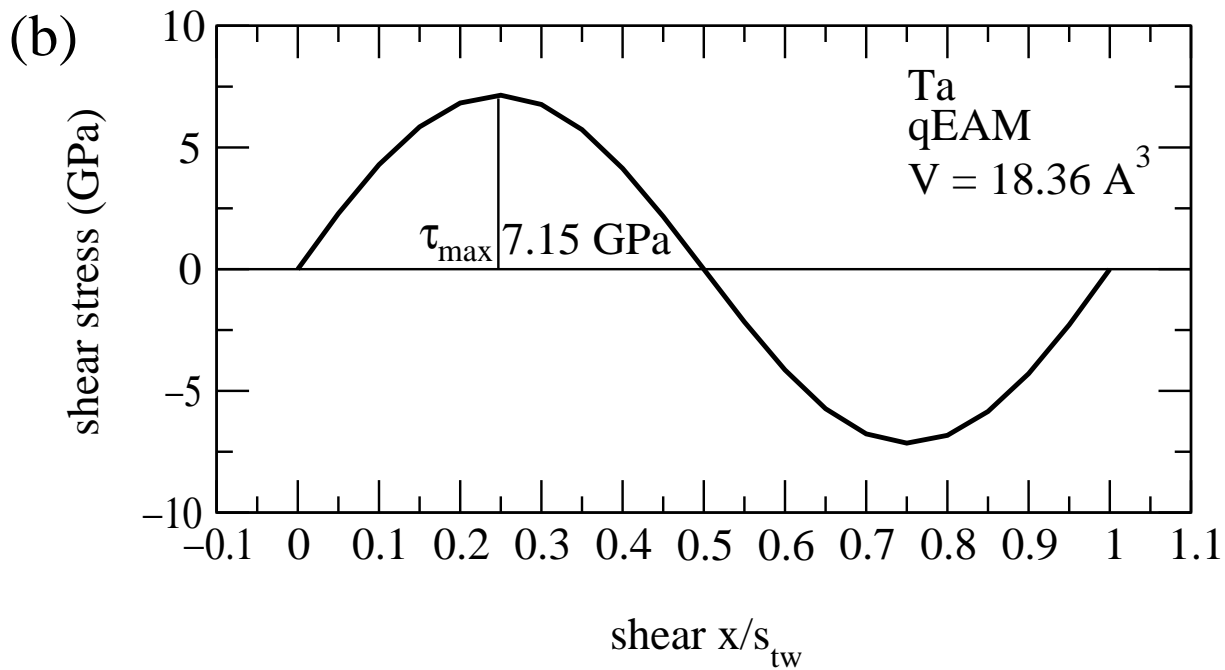
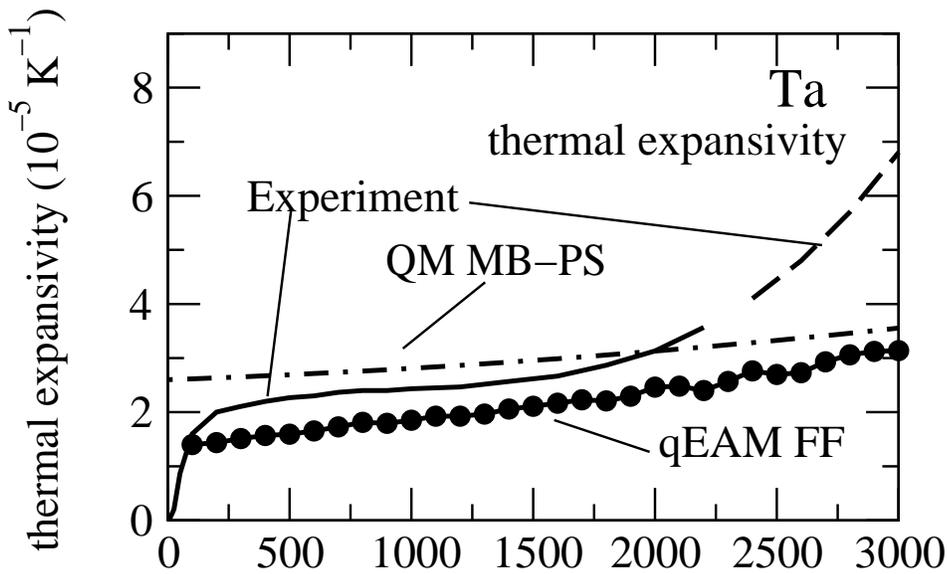
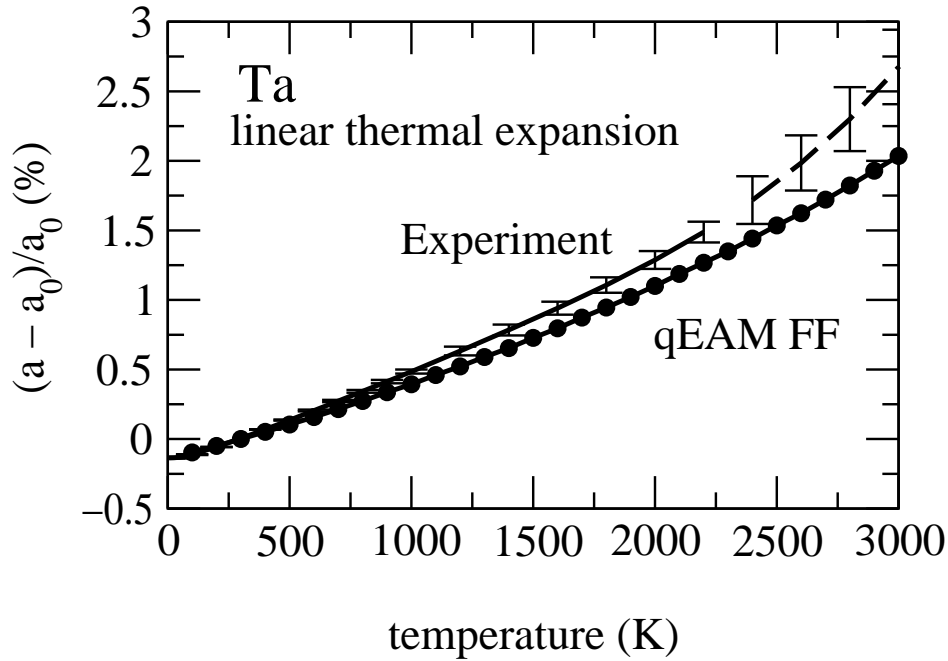
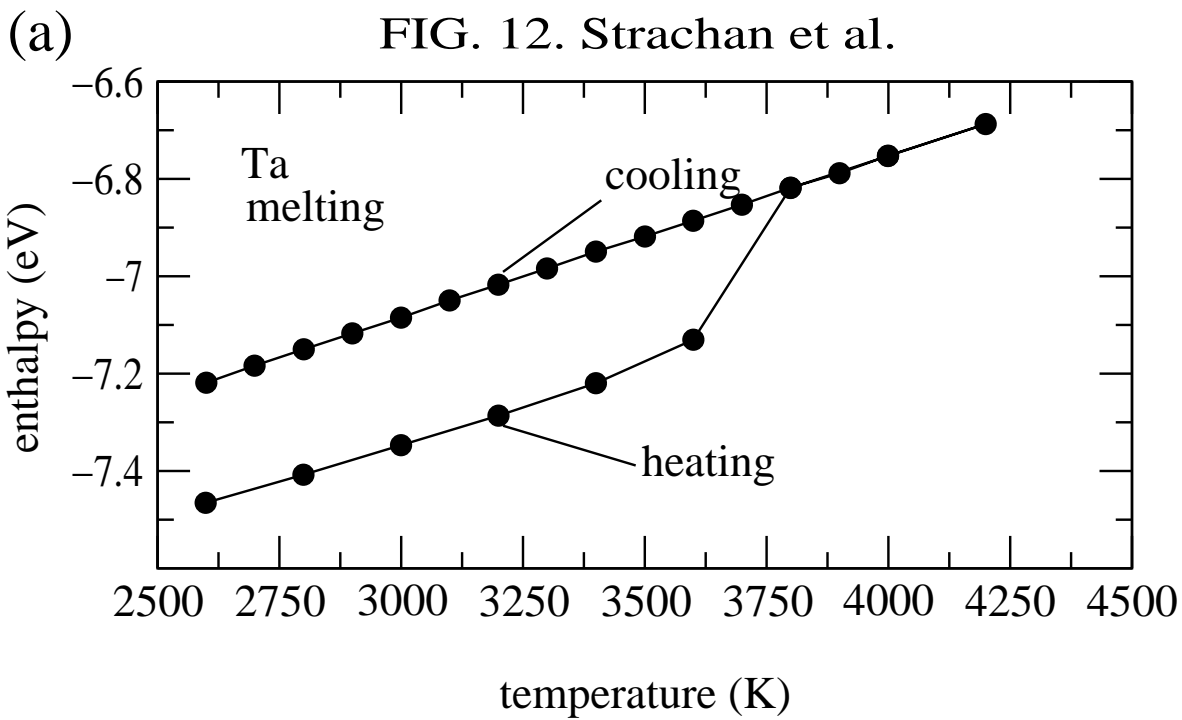
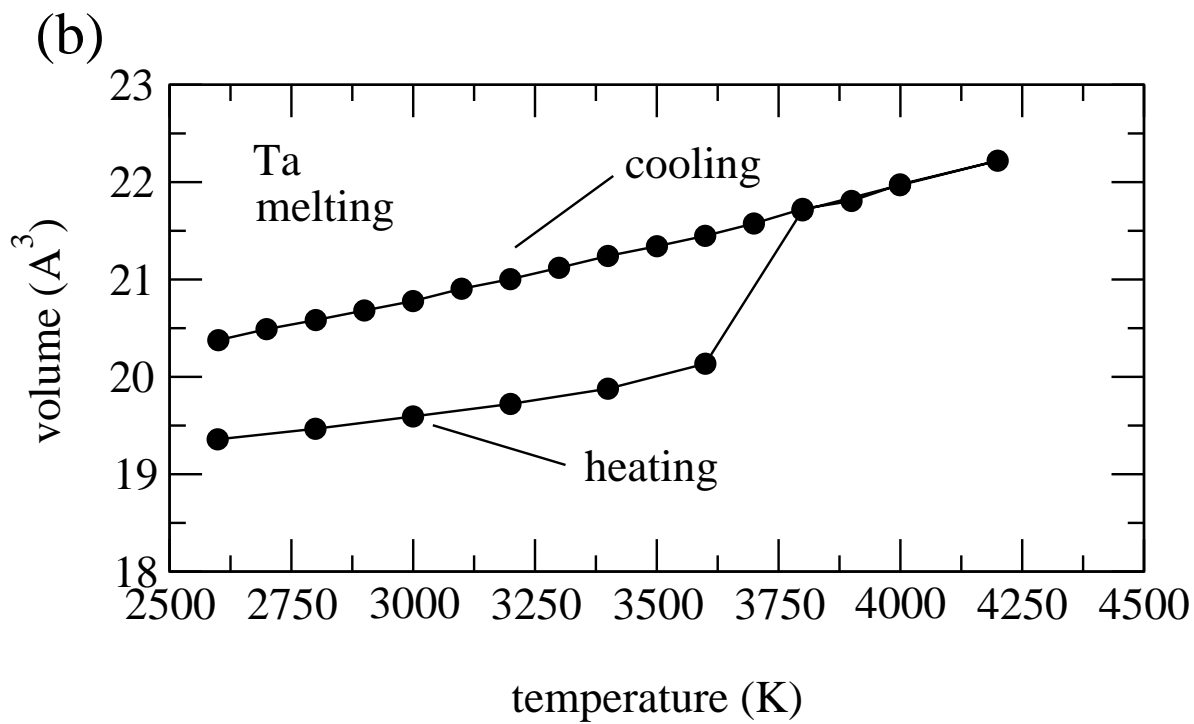
\end{document}